\documentstyle[aps,multicol,pre,epsfig]{revtex}
\begin{document}
\def\sech{\mathop{\rm sech}\nolimits}
\def\csch{\mathop{\rm csch}\nolimits}
\def\span{\mathop{\rm Span}\nolimits}
\def\sn{\mathop{\rm sn}\nolimits}
\def\cn{\mathop{\rm cn}\nolimits}
\def\dn{\mathop{\rm dn}\nolimits}

\begin{title}
{\bf Breather lattice and its stabilization for the modified 
Korteweg-de Vries equation}
\end{title}

\author{P.G. Kevrekidis$^\ast$, Avinash Khare$^\dag$ and A. Saxena$^\ddag$}
\address{$^\ast$Department of Mathematics and Statistics, University of 
Massachusetts, Amherst MA 01003-4515, \\ 
$^\dag$Institute for Physics, 
Bhubaneswar, Orissa 751005, India, \\
$^\ddag$Theoretical Division, Los 
Alamos National Laboratory, Los Alamos, New Mexico 87545.}
\date{\today}
\maketitle

\begin{abstract}
We obtain an exact solution for the breather lattice solution of
the modified Korteweg-de Vries (MKdV) equation.  
Numerical simulation of the breather lattice demonstrates its 
instability due to the breather-breather interaction.  
However, such multi-breather structures can be stabilized through 
the concurrent application of ac driving 
and viscous damping terms.
\end{abstract}

\section{Introduction} 
There are many physical systems where the $1+1$ MKdV equation \cite{mkdv,AS} 
appears, e.g. phonons in anharmonic lattices \cite{ono}, ion acoustic 
solitons \cite{ion} and van Alfv\'en waves in collisionless plasma 
\cite{plasma}, Schottky barrier transmission lines \cite{schot} as well 
as in the models of traffic congestion \cite{traffic}.  A subclass of 
hyperbolic surfaces \cite{schief}, slag-metallic bath interfaces \cite{slag},
curve motion \cite{curve}, meandering ocean jets \cite{jets} and other  
models in fluid mechanics \cite{fluid} are also related to the MKdV 
equation.  Furthermore, it has been shown that the dynamics of thin elastic 
rods can also be reduced to the MKdV equation \cite{mat}.  This equation is 
also of special interest due to its  integrability in the context of 
nonlinear soliton bearing systems \cite{mkdv,AS}.  From a physical 
perspective it is therefore important to examine novel classes of solutions 
of such partial differential equations and their potential relevance in 
this diverse class of applications.

A particularly interesting type of solution is the so called breather 
lattice solution. Breathers are spatially localized and temporally periodic 
solutions which are of significant relevance to localization type 
phenomena in optics, condensed matter physics and biophysics.  For  
a representative set of reviews of the continuously increasing volume 
of work in this area, see e.g. \cite{review}. As can be seen in these 
investigations, typically analytical expressions for breather type 
solutions are unavailable and such solutions have to be traced by means 
of numerical methods. However, in some cases and particularly for 
integrable models, such solutions may exist in closed form. One such 
example is the breather lattice solution of the sine-Gordon equation which 
was presented in explicit analytic form in \cite{mcl}. Such a solution, 
albeit unstable, is very important because it can be stabilized in more 
realistic contexts of driving and damping.  In addition, it is useful in 
extracting the asymptotic breather-breather (exponential) interaction by 
means of energy methods as demonstrated in \cite{ksb}.

It is naturally worthwhile to enquire whether similar extended pattern
solutions are available in closed form in other models. Of particular
interest are models of the MKdV (and KdV) variety not only due to an  
abundance of the corresponding applications but also due to the fundamental
differences of such integrable models with nonlinear Klein-Gordon 
equations, such as sine-Gordon. The latter have Lorentz invariance that 
permits the breathers to be either standing or travelling.  In the former
case such invariances are absent and breathers can only be sustained in a 
travelling wave form.  Furthermore, the dissipation type effects are 
introduced by very different operators in the two cases.  Thus, for 
reasons of physical applicability and mathematical tractability, it
is very important to identify similar solutions of the breather lattice
type in models of the KdV family such as the MKdV and examine their 
stability.  This is the main objective of the present work.
 
Our presentation is structured as follows.  In Sec. II we present the 
explicit MKdV breather lattice solution in terms of elliptic functions, 
retrieve the well known single breather limit and analyze the algebraic
and physical conditions for the existence of such a solution. In Sec.  
III we numerically study the stability of the unperturbed MKdV breather
lattice and display its instability. In Sec. IV, we demonstrate that
such a solution can be stabilized by viscous effects when combined with
appropriate driving to sustain the breathers and finally in Sec. V we 
summarize our findings and present our conclusions.

\section{Breather lattice solution} 
The modified Korteweg-de Vries equation for a field $u(x,t)$ 
$$u_t+6u^2u_x+u_{xxx}=0~, \eqno(1)$$
can be transformed into
$$(1+\phi^2)(\phi_t+\phi_{xxx})+6\phi_x(\phi^2_x-\phi\phi_{xx})=0~, \eqno(2)$$
where
$$u=v_x,~~~\phi=\tan(v/2),~~~and~~~v\rightarrow0~~~as~~|x|\rightarrow\infty~. 
\eqno(3)$$

The single breather solution of MKdV is known to be \cite{mkdv}
$$u(x,t)=-2\frac{\partial}{\partial x}\tan^{-1}\left(\frac{c\sin(ax+bt+a_0)}
{a\cosh(cx+dt+c_0)}\right)~, \eqno(4) $$
with
$$b=a(a^2-3c^2)~, ~~~d=c(3a^2-c^2)~, \eqno(5)$$
and $a_0$, $c_0$ are arbitrary constants.

Let us now try to obtain the breather lattice solution of the MKdV 
equation.  To that end, we start with the ansatz
$$u(x,t)=-2\frac{\partial}{\partial x}\tan^{-1}\phi(x,t)~, \eqno(6)$$ with
$$\phi(x,t)=\alpha \sn(ax+bt+a_0,k)\dn(cx+dt+c_0,m)~, \eqno(7)$$
where $\sn(x,k)$ and $\dn(x,m)$ are Jacobi elliptic functions with modulus 
$k$ and $m$, respectively.  This ansatz is inspired by the derivative 
relationship between a single breather solution of the sine-Gordon 
equation and that of MKdV [Eq. (4)] as well as by the functional similarity 
of the sine-Gordon breather lattice solution \cite{mcl,ksb}. Substituting 
the ansatz (7) in Eq. (2) 
and upon lengthy algebraic manipulations, we find that Eq. (7) is indeed the 
MKdV 
breather lattice solution, provided that
$$a^4k=c^4(1-m)~,~~\alpha =-\frac{c}{a}~, \eqno(8)$$
$$b=a[a^2(1+k)-3c^2(2-m)]~, ~~d=c[3a^2(1+k)-(2-m)c^2]~. \eqno(9)$$
As expected, in the limit $m \rightarrow 1, k \rightarrow 0$, the breather
lattice solution (7) reduces to the single breather solution (4) and the
relations between $c$ and $d$ as well as  between $a$ and $b$ reduce 
to the ones given by Eq. (5).

On physical grounds (i.e., to have solutions with a definite spatial 
periodicity), it is natural to demand that the periods of the $\sn(x,k)$ 
and $\dn(x,m)$ functions must be spatially commensurate, i.e., in addition 
to conditions (8) and (9), we must also demand that \cite{footnote} 
$$4K(k)/a=2K(m)/c~, \eqno(10)$$
where K(k) is the complete elliptic integral of the first kind.  Note, 
however, that since the MKdV breather is always moving, it need not
have temporal commensurability.  Combining conditions (8) and (10)
yields
$$16kK^4(k)=(1-m)K^4(m)~, \eqno(11)$$
implying thereby that $m$ and $k$ are not independent. For example, note
that as expected with $m \rightarrow 1$, $k \rightarrow 0$ the breather
lattice solution reduces to the single breather solution (4). However, as
$m \rightarrow 0$, $\dn(x,m)=1$ and then it is easily shown that an exact
(nonlinear) travelling wave solution is given by
$$\phi(x,t)=\sqrt{k}\sn(ax+a^3t[1+k-6\sqrt{k}]+a_0,k)~. \eqno(12)$$

\begin{figure}
\centering
{\epsfig{file=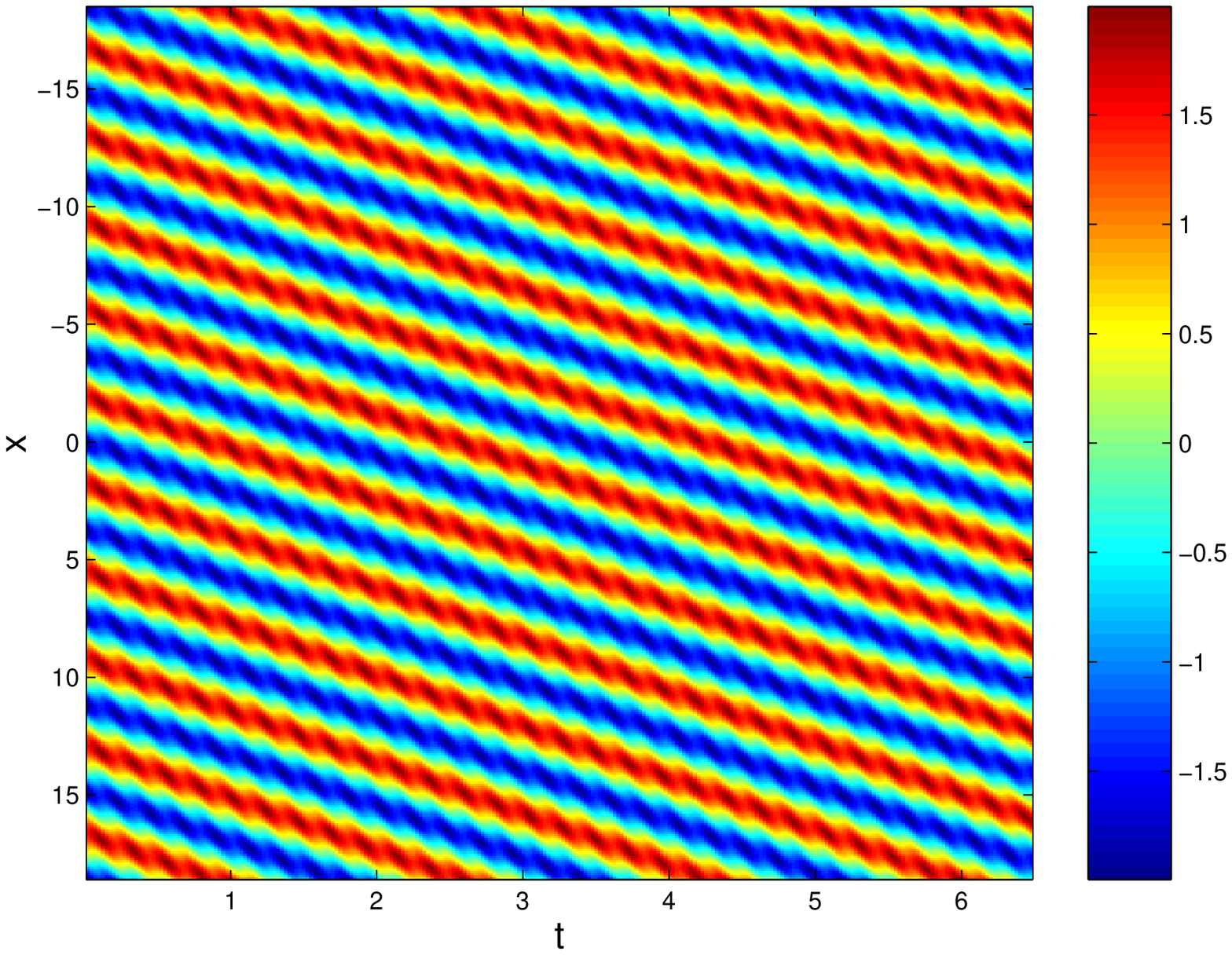, width=6.cm,angle=0, clip=}}
{\epsfig{file=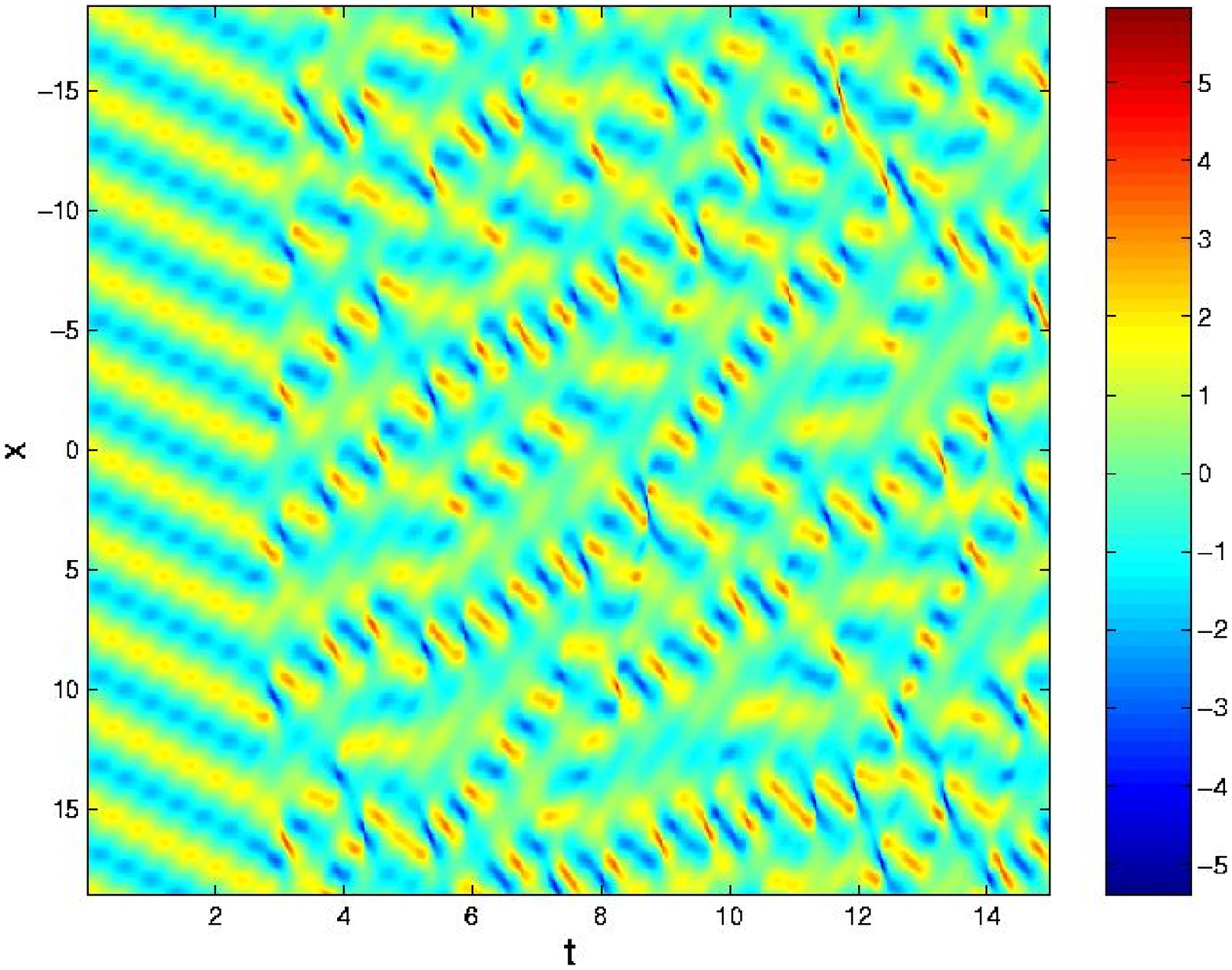, width=6.cm,angle=0, clip=}}
{\epsfig{file=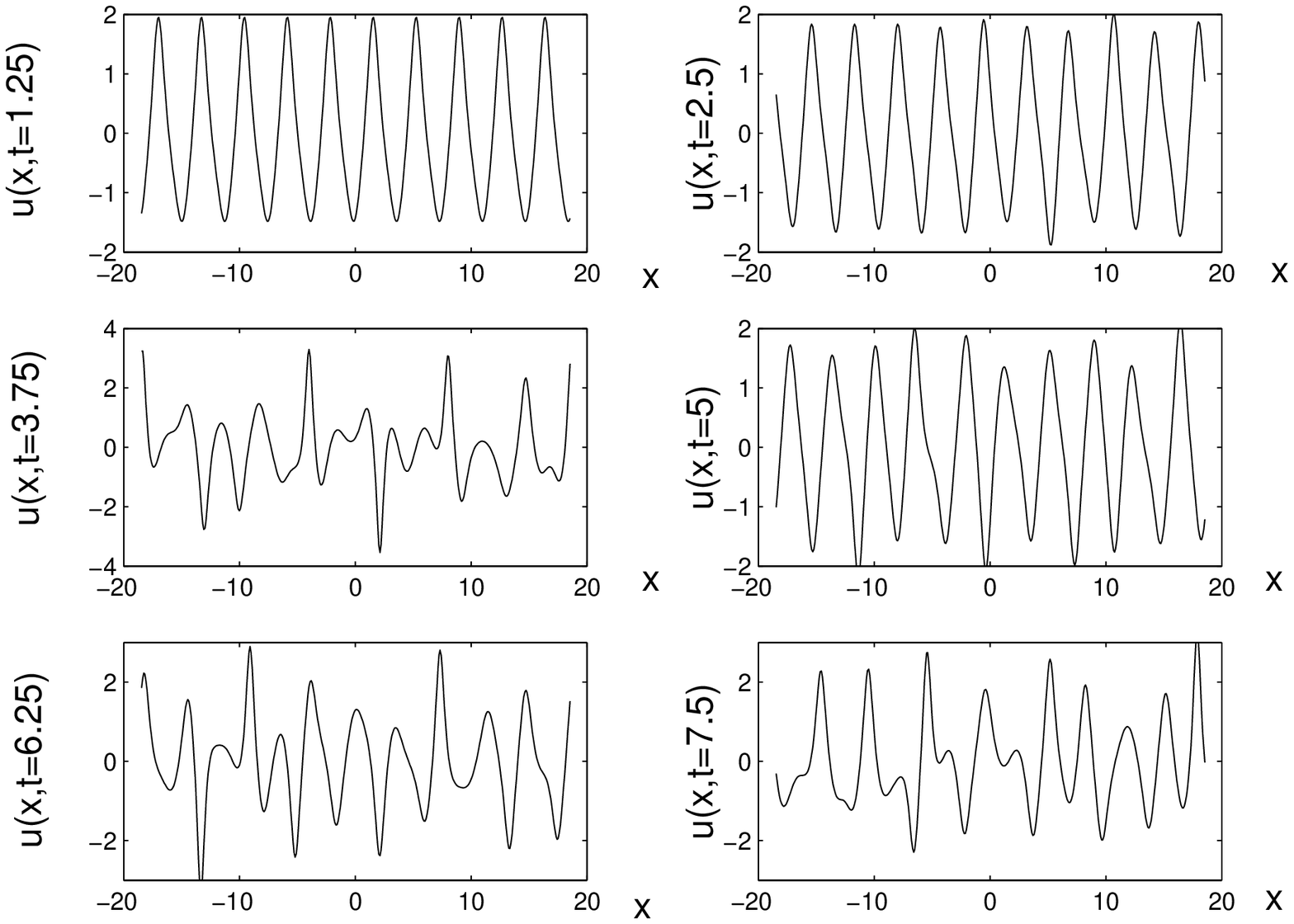, width=6.cm,angle=0, clip=}}
\caption{Pictorial representation of the exact MKdV breather lattice 
solution in a space-time $(x,t)$ contour plot for $c=1$, $m=0.5$ (top 
left panel).  The top right panel shows the time evolution of the same
solution under the dynamics of Eq. (13) with $\Delta t=0.0005$ and $h 
\approx 0.0927$.  These contour plots have dimensionless units for $x$ 
and $t$.  The bottom panel shows the spatial profiles $u(x,t_0)$ of the
solution for various times $t_0$ before and after the instability develops.}
\label{afig1}
\end{figure}

Summarizing, since there are four relations among the six parameters
$a,b,c,d,k,m$, we have obtained a two-parameter family of breather
lattice solutions. A plot of the (exact) breather lattice of Eq. (6)
for $m=0.5$, 
$c=1$ is given in the top left panel of Fig. \ref{afig1}.

\section{Numerical methods and evolution of the breather lattice} 

Based on the breather lattice simulations for the sine-Gordon case 
\cite{ksb} it may be natural to expect that the MKdV breather lattice 
configuration is also unstable. In the numerical simulation of the MKdV 
problem, we have found that the direct center-difference discretization 
does a poor job in adequately following the MKdV equation (and in 
conserving the corresponding integrals of motion). While one can also use 
the integrable scheme of Ablowitz-Ladik (see e.g., \cite{AS,AL,invp} and 
references therein), we have followed a different path here in spatially 
discretizing the partial differential equation and following the integrable 
discretization scheme of \cite{ohta} for KdV and adapting it to the case 
of the MKdV.  In particular, the spatially discrete version of our 
equation reads (with lattice spacing $h$):
$$\dot{u}_n=-\frac{1}{2 h^3} \left( u_{n+2}- 2 u_{n+1} 
+ 2 u_{n-1} - u_{n-2} \right) - \frac{1}{3 h}
\left[ u_{n+1}^2 \left( u_{n+2} + u_{n+1} + u_n \right)
      -u_{n-1}^2 \left( u_{n} + u_{n-1} + u_{n-2} \right) \right] . 
\eqno(13) $$ 
The time integration has been performed by means of a 4th order 
Runge-Kutta scheme.  We used periodic boundary conditions and the 
initial condition contained an exact breather lattice configuration, 
matching the periodicity of the finite domain. Hence, the only 
perturbation to the exact solution came from the numerical discretization 
of the problem. It should also be noted that in the results mentioned 
below, the accuracy of the numerical method was monitored by probing the 
conservation of two quantities, $\sum_n u_n$ and $\sum_n u_n^2$, which 
emulate the discrete analogs of the mass and the momentum, respectively. 
Typically the former is conserved (at worst) to 1 part in $10^7$, while 
the latter to 1 part in $10^3$.

We found that, as can be seen in the top right panel of Fig. \ref{afig1}, 
the numerical discretization perturbation grows and eventually destroys 
the breather lattice configuration in all the cases considered. Various 
snapshots of the solution $u(x,t)$ are depicted in the bottom panel of 
Fig. \ref{afig1}. 

\begin{figure}
\centering
{\epsfig{file=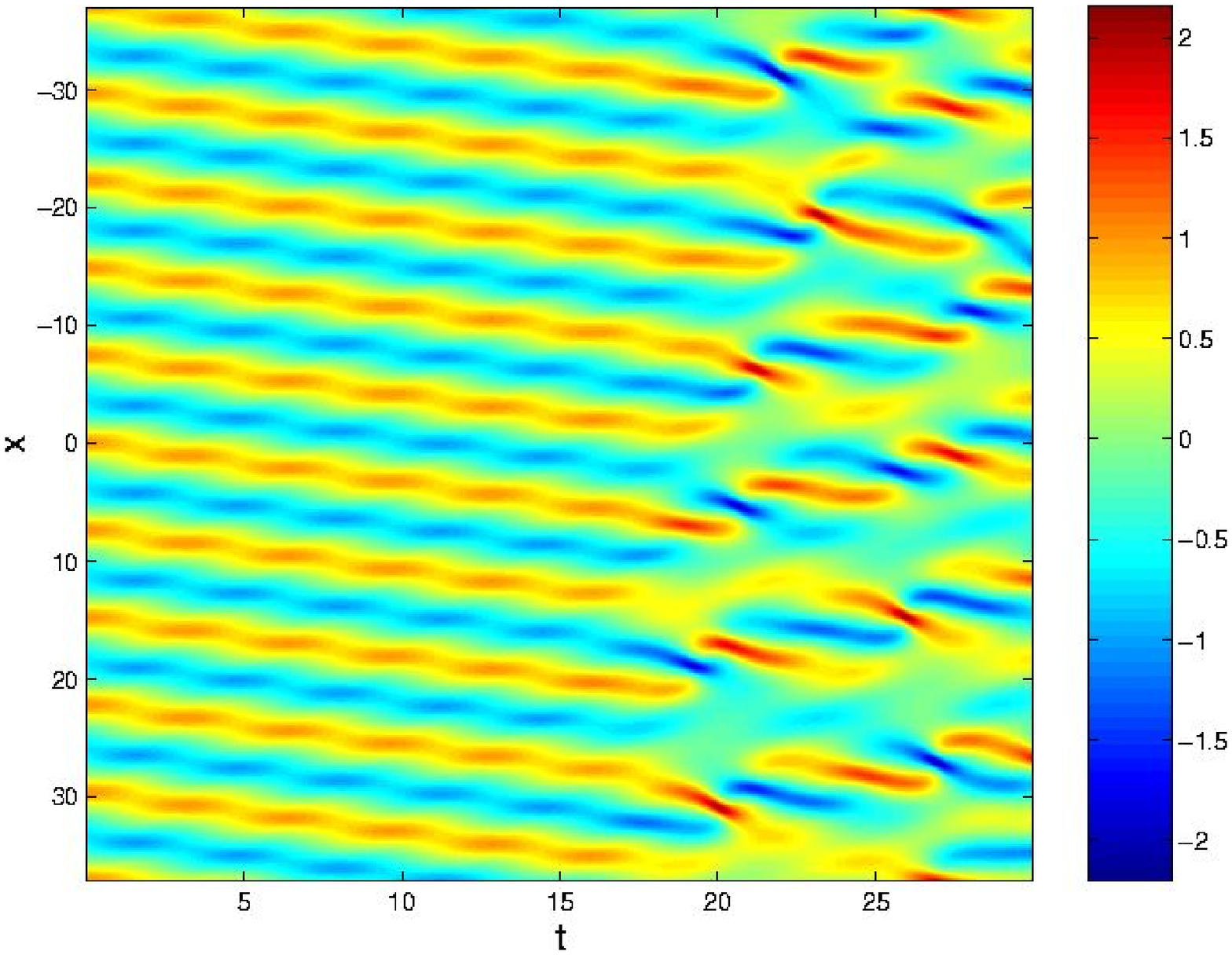, width=6.cm,angle=0, clip=}}
{\epsfig{file=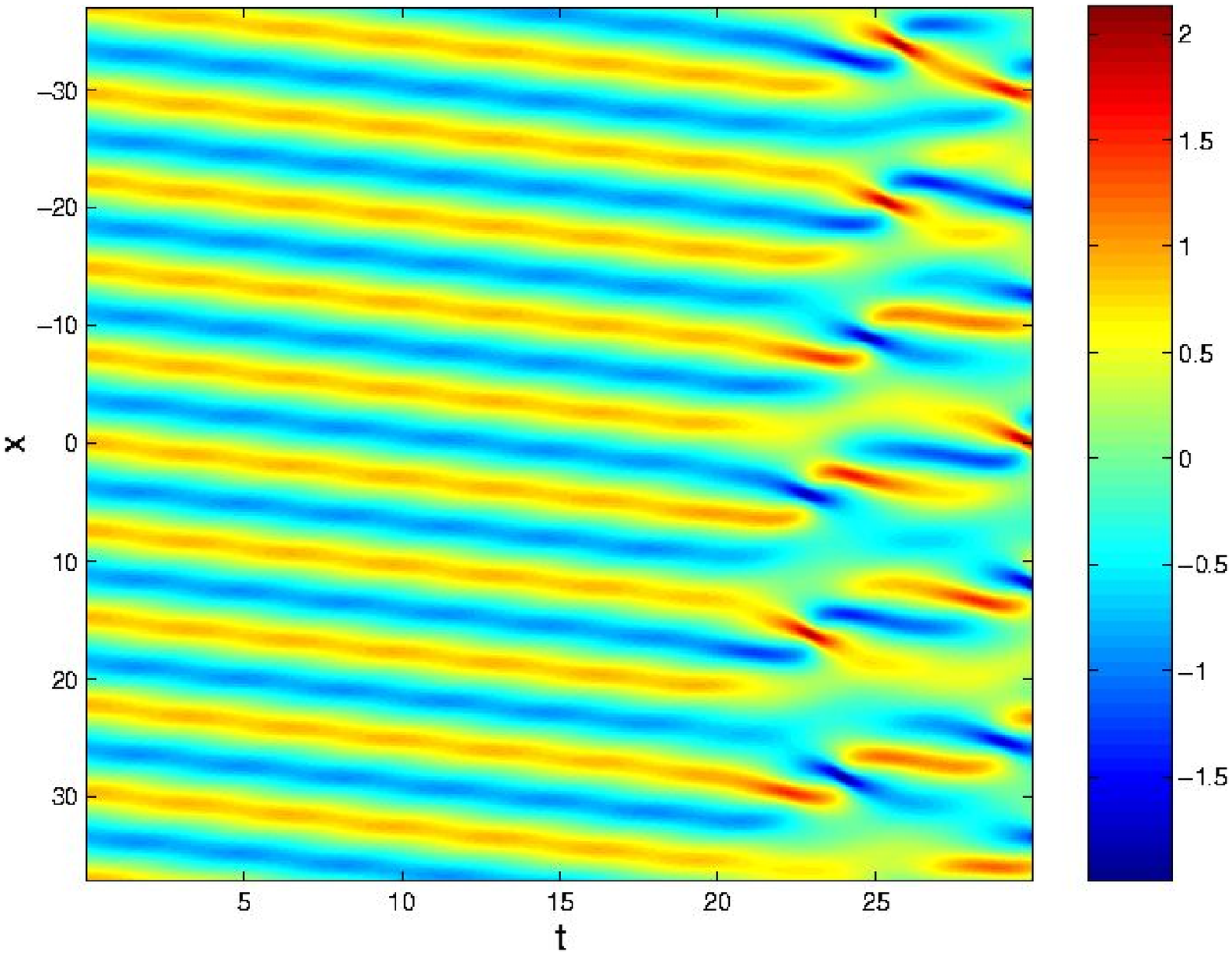, width=6.cm,angle=0, clip=}}
{\epsfig{file=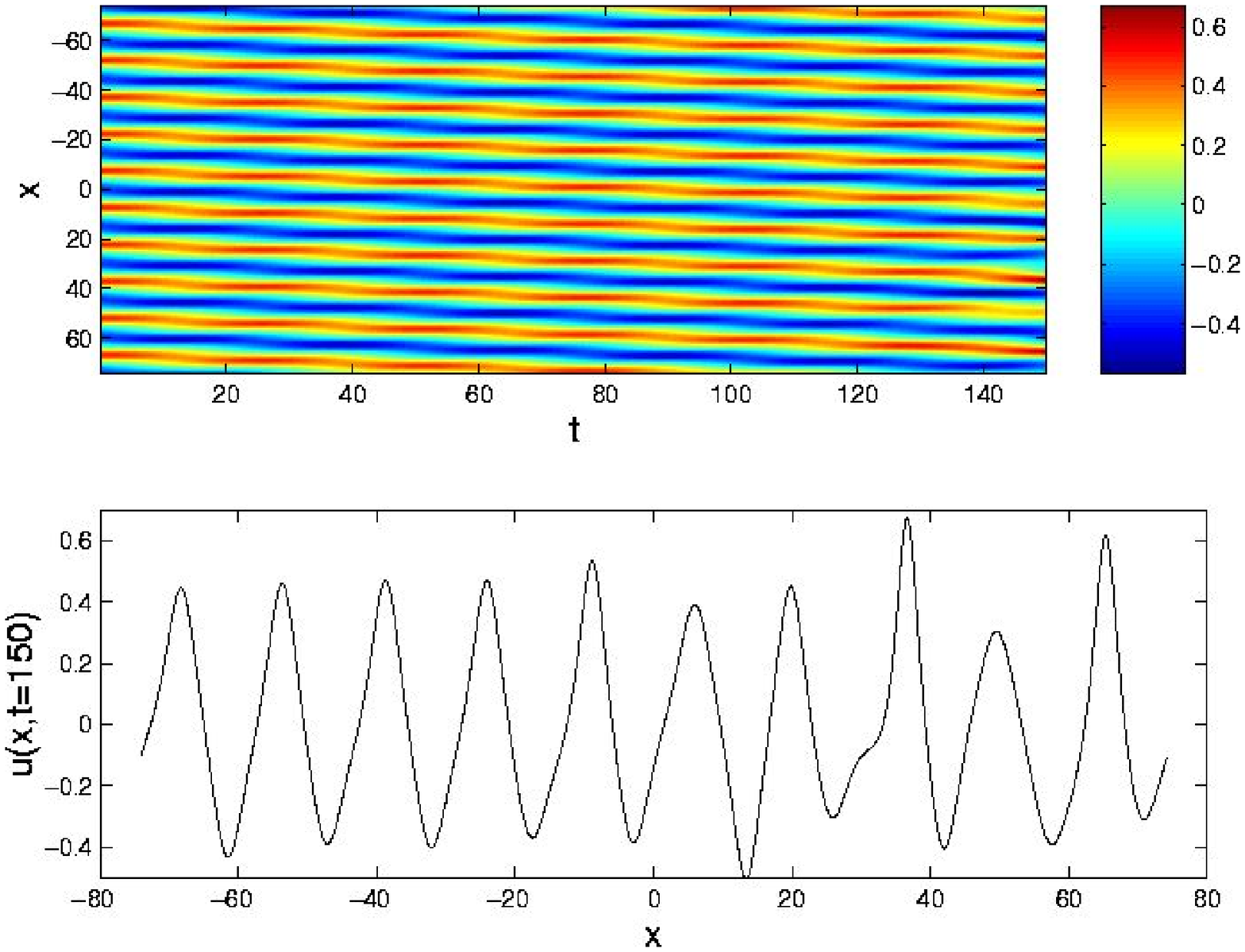, width=6.cm,angle=0, clip=}}
\caption{Same as in the top panels of Fig. 1, but now in the top left 
panel $c=0.5$ and $m=0.5$, in the top right panel $c=0.4546$ and 
$m=0.25$, and in the bottom panel $c=0.25$ and $m=0.5$.  In the latter 
case the snapshot of $t=150$ is also shown to indicate the onset of the 
instability.  The contour plots have dimensionless units for $x$ and $t$.}
\label{afig2}
\end{figure}

To further understand the instability, we performed runs for 
different values of the two relevant parameters of the solution,
namely $c$ and $m$. Three typical cases are shown in Fig. \ref{afig2}. 
The particular numerical experiments are chosen to illustrate the 
characteristic dependences of the instability.  We expect from the 
experience of other nonlinear wave equations with interacting breather 
structure (see e.g., \cite{ksb} for sine-Gordon and \cite{kkm} for 
nonlinear Schr{\"o}dinger type models) that the instability is caused 
by the interaction between the breathers, which is exponential in their 
separation. From Eq. (10), the separation between the breathers is 
given by $S=2 K(m)/c$. The top panels of Fig. \ref{afig2} correspond 
to two cases with different $c$ and $m$ but with the same $S \equiv S_0$. 
Clearly 
the instability develops at very similar times and verifies the 
dependence of the growth rate on the inter-breather separation $S_0$. 
The bottom panel shows a case with $S=2 S_0$. Notice that these typical 
results have been verified by additional numerical experiments. 
Furthermore, the ratio of the breather separations in the cases of the 
top right panel of Fig. \ref{afig1}, top panels of Fig. \ref{afig2} and 
the bottom panel of Fig. \ref{afig2} is 1:2:4 while the corresponding 
(approximate) instability onset times have a ratio of 2.5:16:150, clearly 
hinting an exponential dependence of the instability onset on $S$. 

\section{Stabilization}

\begin{figure}
\centering
{\epsfig{file=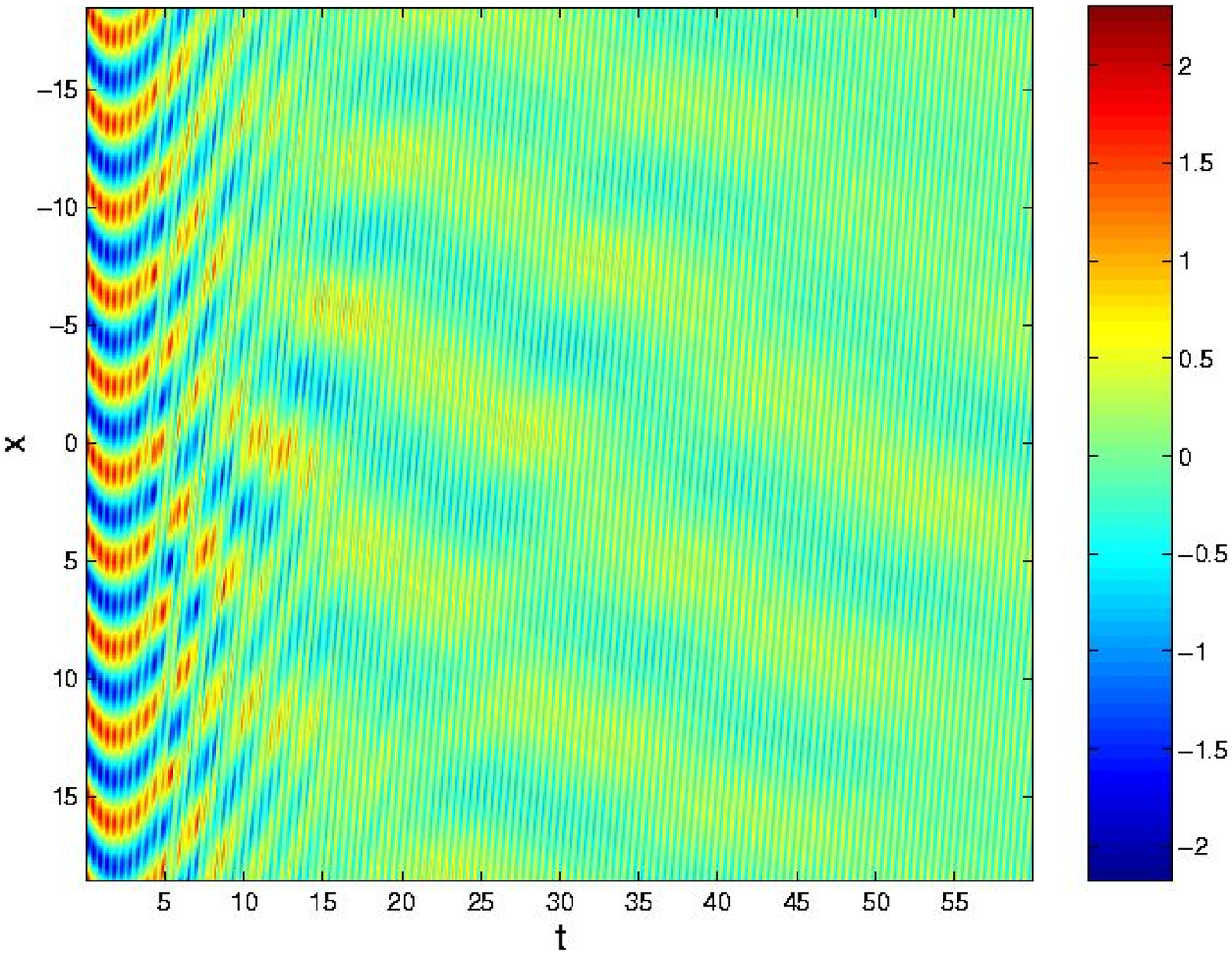, width=6.cm,angle=0, clip=}}
{\epsfig{file=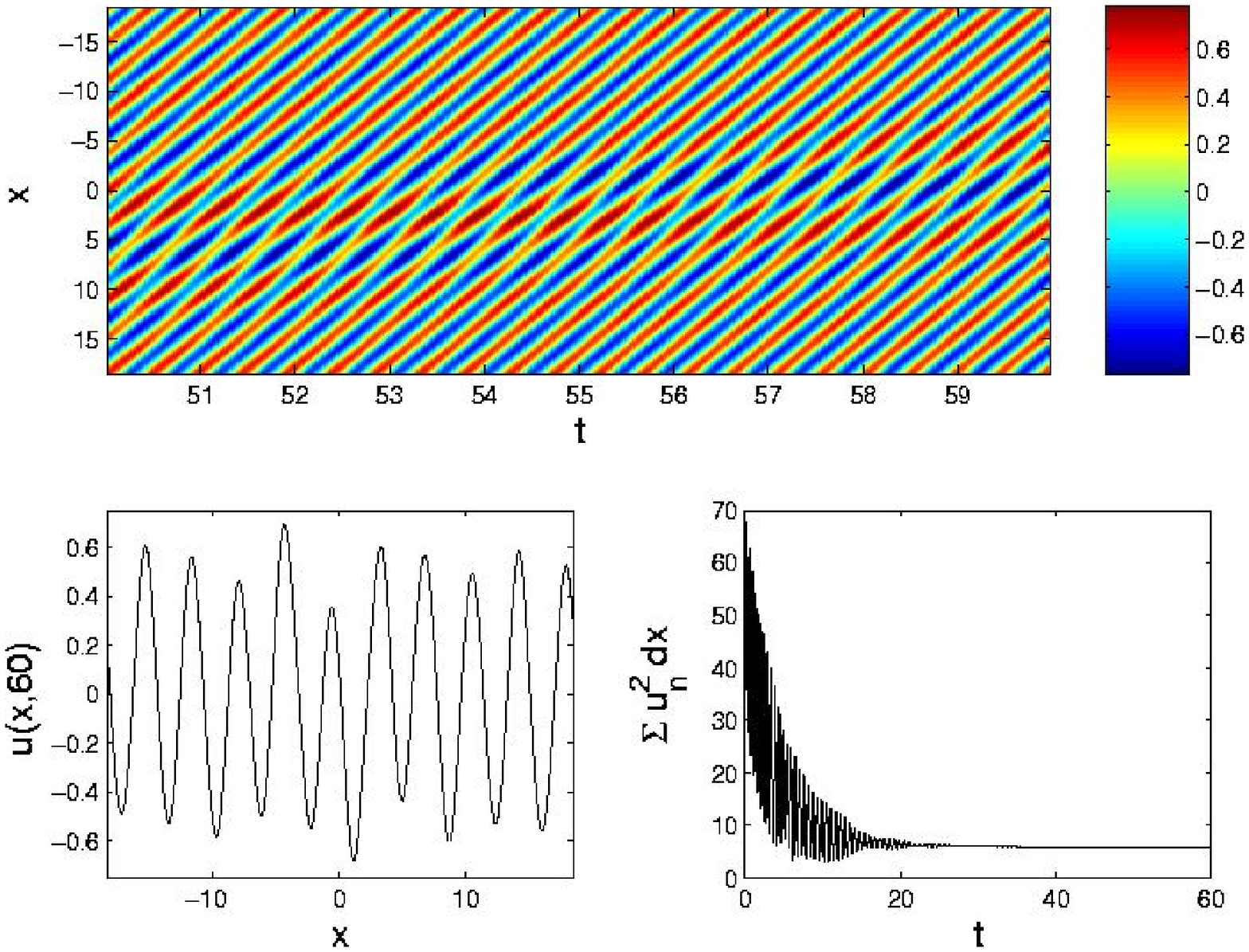, width=6.cm,angle=0, clip=}}
\caption{The driven-damped MKdV stabilization of the breather lattice 
with $c=1$, $m=0.5$, $\beta=F_0=5$. The left panel shows the space-time
contour plot. The right panel top subplot shows a detail of the left 
plot to indicate the stabilization of the configuration. These contour 
plots have dimensionless units for $x$ and $t$.  The bottom
left subplot shows the spatial profile of the solution at $t=60$, while 
the bottom right subplot depicts the relaxational time evolution of a 
global property such as the discrete analog of the continuum momentum.}
\label{afig3}
\end{figure}

While the results of the previous section indicate that very long-lived
breather lattice configurations can be achieved by appropriate parameter
selection, it is natural to enquire whether by mechanisms of ac 
driving and damping it is possible to fully stabilize such configurations.
We have thus examined the following driven-damped MKdV equation  
$$u_t + u_{xxx}+ 6 u^2 u_x = \beta u_{xx} + 
F_0\sin\left(\frac{\pi}{K(m)}(c x+\omega t)\right) , \eqno(14)  
$$ 
where the ``viscosity'' coefficient was fixed to $\beta=5$ while $F_0$ 
was varied. Notice that the periodicity of the driver was chosen to 
match one of the unperturbed breather lattice configurations. For small 
values of $F_0$, the viscosity damps the breather amplitude. However, 
for sufficiently large driving amplitudes (such as the one used in Fig. 
\ref{afig3}) the driver can lead to the stabilization of an asymmetric 
lattice configuration.  We also note that the ac drive was motivated by 
earlier studies, e.g. \cite{malkov} (and the references therein). In 
addition, we point out that lattice configurations propagating in the 
opposite direction can be stabilized if the velocity of the driver is 
reversed (results not shown here).

We note that the value of the viscosity coefficient $\beta=5$ and the 
amplitude of the driver $F_0=5$ used in Fig. 3 to stabilize the breather 
lattice are relatively large.  In this sense the stabilized (numerical) 
breather lattice in Fig. 3 is different from the exact solution of the 
MKdV equation given in Eqs. (6) and (7).  Thus the physical origin of 
the stabilized solution may be different from the original solution of 
the MKdV equation.

We should also remark that 
variation of the value of $\beta$ will not significantly modify the
results. I.e., even for much smaller values of $\beta$ the multi-breather
configuration is destroyed at a finite time and the resulting profile
does not closely resemble the initial condition. However, the relaxation
time to the final state depends considerably on the exact value of 
the viscosity coefficient and is longer for smaller $\beta$.

\section{Conclusion}

Inspired by the exact breather lattice solution of the sine-Gordon
equation \cite{mcl,ksb} we used an ansatz to find a corresponding
solution of the modified Korteweg-de Vries equation.  We determined
the conditions under which the ansatz becomes an exact solution of
MKdV and showed  how it degenerates to the single MKdV breather
solution in the appropriate limit.  We then used this exact expression
to derive additional lattices of nonlinear travelling waves. The MKdV
breather lattice is a genuinely propagating solution in constrast to the
sine-Gordon solution which can be static.  Our numerical experiments
(by means of a novel numerical scheme) indicated that the MKdV breather
lattice solution is unstable; however, it can be stabilized by
inclusion of damping  and ac driving.  The results presented here may
be relevant to numerous physical phenomena such as jamming in traffic
flow \cite{traffic}, fluid dynamics \cite{fluid} and collisionless
plasmas \cite{plasma}.

This research is supported in part by the U.S. Department of Energy 
under contract W-7405-ENG-36 and in part by NSF under DMS-0204585.


\begin{references}

\bibitem{mkdv} {\it Solitons: an introduction}, P.G. Drazin and R.S. 
Johnson (Cambridge University Press, Cambridge, U.K., 1989).  

\bibitem{AS} M.J. Ablowitz and H. Segur,
\newblock Solitons and the Inverse Scattering Transform
\newblock (SIAM, Philadelphia, 1981).

\bibitem{ono} H. Ono, J. Phys. Soc. Jpn. {\bf 61}, 4336 (1992). 

\bibitem{ion} K. E. Lonngren, Optical and Quantum Electronics, 
{\bf 30}, 615 (1998). 

\bibitem{plasma} A. H. Khater, O. H. El-Kakaawy, and D. K. Callebaut, 
Phys. Script. {\bf 58}, 545 (1998).  

\bibitem{schot} V. Ziegler, J. Dinkel, C. Setzer, and K. E. Lonngren, 
Chaos, Solitons and Fractals {\bf 12}, 1719 (2001).  
 
\bibitem{traffic} T. S. Komatsu and S. I. Sasa, Phys. Rev. E {\bf 52},
5574 (1995); T. Nagatani, Physica A {\bf 265}, 297 (1999). 

\bibitem{schief} W. K. Schief, Nonlinearity {\bf 8}, 1 (1995). 

\bibitem{slag} M. Agop and V. Cojocaru,
Mater. Trans. JIM {\bf 39}, 668 (1998). 

\bibitem{curve} J. Langer and R. Perline, Phys. Lett. A {\bf 239}, 
36 (1998); K. S. Chou and C. Z. Qu, Physica D {\bf 162}, 9 (2002). 

\bibitem{jets} E. A. Ralph and L. Pratt, J. Nonlin. Sci. {\bf 4}, 
355 (1994). 

\bibitem{fluid} M.A. Helal,
Chaos, Solitons and Fractals {\bf 13}, 1917 (2002).

\bibitem{mat} S. Matsutani and H. Tsuru, J. Phys. Soc. Jpn. {\bf 60}
(1991) 3640.

\bibitem{review} S. Aubry, \newblock Physica {\bf 103D}, 201 (1997);
S. Flach and C.R. Willis, \newblock Phys. Rep. {\bf 295}, 181 (1998);
Physica {\bf 119D}, (1999), special volume edited by S. Flach and R.S. 
MacKay; P.G. Kevrekidis, K.{\O}. Rasmussen and A.R. Bishop, Int. J. 
Mod. Phys. B {\bf 15}, 2833 (2001); focus issue edited by Yu. S. 
Kivshar and S. Flach, Chaos {\bf 13}, 586-799 (2003). 

\bibitem{mcl} R. McLachlan, Math. Intelligencer {\bf 16}, 31 (1994).

\bibitem{ksb} P. G. Kevrekidis, A. Saxena, and A. R. Bishop, Phys. Rev. 
E {\bf 64}, 026613 (2001). 

\bibitem{footnote} We have also examined ``weakly commensurate'' cases 
where $ 4 p K(k)/a=2 q K(m)/c$ with similar results.  Here $p$ and $q$ 
denote relative primes.  For incommensurate cases, the boundary 
periodicity cannot be imposed and typically we found that this creates 
an immediate instability-inducing perturbation to the configuration. 
 

\bibitem{AL} M.J. Ablowitz and J.F. Ladik, \newblock J. Math. Phys. 
{\bf 16}, 598 (1975); {\it ibid.} {\bf 17}, 1011 (1976).

\bibitem{invp} A. Mukaihira and Y. Nakamura,
\newblock Inverse Probl. {\bf 16}, 413 (2000).

\bibitem{ohta} Y. Ohta and R. Hirota,
\newblock J. Phys. Soc. Jpn. {\bf 60}, 2095 (1991).

\bibitem{kkm} T. Kapitula, P.G. Kevrekidis, and B.A. Malomed,
\newblock Phys. Rev. E, {\bf 63}, 036604 (2001).

\bibitem{malkov} M.A. Malkov,
Physica D {\bf 95}, 62 (1996).

 
\end{references}
\end{document}